\documentstyle[aps,preprint]{revtex}

\input prepictex
\input pictex
\input postpictex

\newcommand{\TeV}{\,{\rm TeV}}
\newcommand{\GeV}{\,{\rm GeV}}
\newcommand{\MeV}{\,{\rm MeV}}
\newcommand{\keV}{\,{\rm keV}}
\newcommand{\eV}{\,{\rm eV}}

\def\npb#1#2#3{Nucl.\ Phys.\ {B#1}, #2 (#3)}
\def\prd#1#2#3{Phys.\ Rev.\ {D#1}, #2 (#3)}
\def\prl#1#2#3{Phys.\ Rev.\ Lett.\ {#1}, #2 (#3)}
\def\plb#1#2#3{Phys.\ Lett.\ {B#1}, #2 (#3)}
\def\prt#1#2#3{Phys.\ Rep.\ {#1}, #2 (#3)}

\begin{document}

\preprint{\begin{tabular}{r}
KAIST--TH 97/14\\[-3mm]
KIAS-P 97014 \\[-3mm]
hep-ph/9801280
\end{tabular}}


\title{Cosmology of Light Moduli}
\author{Kiwoon Choi$^1$, Eung Jin Chun$^2$ and Hang Bae Kim$^1$}
\address{$^1$Department of Physics,
	Korea Advanced Institute of Science and Technology\\
	Taejon 305-701, Korea\\
	$^2$ Korea Institute for Advanced Study\\
	207-43 Cheongryangri-dong, Dongdaemun-gu, Seoul 130-012,  Korea
}
\maketitle
\begin{abstract} 
In string/$M$-theory with a large compactification radius,
some axion-like moduli can be much lighter than the gravitino.     
Generic  moduli in gauge-mediated supersymmetry breaking models also 
have a mass far below the weak scale.
Motivated by these, we examine the cosmological implications
of  light moduli for the mass range from the weak scale
to an extremely small scale of ${\cal O}(10^{-26})$ eV,
and  obtain an upper bound on the initial  moduli misalignment
for both cases with and without a late entropy production.
\end{abstract}

\section{Introduction}

A very weakly-interacting light scalar field $\phi$ can be cosmologically
troublesome.  In the early universe with the Hubble expansion
parameter  $H$ much bigger
than the scalar boson mass $m_{\phi}$, this scalar field would take an
initial value $\delta \phi$ significantly different from its vacuum value at
present.  Due to this misalignment, $\phi$ starts to coherently oscillate
at a later time and subsequently its energy
density behaves as a non-relativistic matter.
Depending upon its lifetime, such a coherently oscillating scalar field 
may spoil the big bang nucleosynthesis or produce too much gamma- or
$X$-rays through its late-time decay, or may overclose the universe.

The most well-known and first studied example of such a light scalar boson 
is the invisible axion which has been introduced to solve the strong CP
problem \cite{kim,coldaxion}.  The mass and the typical misalignment of the
invisible axion are determined by a single unknown parameter, the axion decay
constant $f_a$ which corresponds to the scale of spontaneous
$U(1)_{PQ}$-breaking, as $\delta a\approx f_a$ and
$m_a\approx f_{\pi}m_{\pi}/f_a$ where $f_{\pi}$ and $m_{\pi}$
are the pion decay constant and the  pion mass, respectively.
If there is no entropy production after the QCD phase transition, 
the requirement that the relic axion mass density does not  overclose
the universe leads to the famous constraint $f_a\lesssim 10^{12}$ GeV.
It was later realized that generic hidden sector supergravity model predicts
a cosmologically troublesome light scalar field,  the Polonyi field,
whose  typical misalignment is  given by  
the Planck scale $M_P=1/\sqrt{8\pi G_N}\approx 2\times10^{18}$ GeV
\cite{polonyi}.
Again in the absence of a late time entropy production after the Polonyi
field starts to oscillate, it would completely spoil the successful
big-bang nucleosynthesis unless the Polonyi mass $m_{\phi}\gtrsim40\TeV$.

This Polonyi problem has been recently revived in the context of string
theory \cite{moduli}.  The (approximately) degenerate string vacua are
described by moduli fields whose typical misalignments are given by
either the string scale or the compactification scale which is close to $M_P$.
In the previous studies, the moduli masses were assumed to be of order the
gravitino mass $m_{3/2}$ which would be of order of the weak scale if the
supersymmetry breaking is transmitted to the supersymmetric standard model
sector by gravitational  interactions.  However due to approximate
non-linear global $U(1)$ symmetries in string/$M$-theory, some of the
axion-like  moduli in string theory can be much lighter than the gravitino
\cite{bank}.  In particular, in models with a large
compactification radius, the mass of such light axion-like moduli behaves as
$m_{\phi}\approx e^{-\pi {\rm Re}({\cal T})}m_{3/2}$ where ${\cal T}$ 
is the overall modulus field
whose value corresponds to the compactification radius-squared in the 
heterotic string length unit.  As we will discuss in section II, 
${\rm Re}({\cal T})$ can be
as large as ${\cal O}({1\over \alpha_{GUT}})$ in the $M$-theory limit, implying
that $m_{\phi}$ can be {\it extremely} smaller than $m_{3/2}$,
for instance as small as $10^{-34}m_{3/2}$.

Although not as dramatic as 
the axion-like moduli in string/$M$-theory,
generic moduli   in models with a gauge-mediated supersymmetry
breaking can also  be much lighter than the weak scale \cite{murayama}. 
In gauge-mediated models,
the supersymmetry breaking scale
$\Lambda$  is related to the weak scale $M_W$ by some loop
suppression factor:
$M_W=({\alpha\over\pi})^l \Lambda$ where the integer $l$ counts
the number of loops involved in 
transmitting supersymmetry breaking
to the supersymmetric standard model sector.
Then for a reasonable value of $l$, the
moduli mass $m_{\phi}\approx \Lambda^2/M_P$
would be in the range
far below the weak scale.

Motivated by the  above  observations, in this paper we wish to study some
cosmological aspects of a generic light modulus  $\phi$ with an arbitrary
mass in the range from the weak scale  to an extremely small mass scale
of ${\cal O}(10^{-26})$ eV. 
It turns out that no meaningful
cosmological bound is obtained for the moduli mass below $10^{-26}$ eV.
In fact, moduli cosmology in gauge-mediated supersymmetry
breaking models has been discussed recently \cite{yana}, but only for
a rather narrow mass range $m_{\phi}=10\keV-1\GeV$.
The organization of this paper goes as follows.
In section II, we discuss in more detail the masses of the light axion-like
moduli in string/$M$-theory to make our motivation more clear.
In section III, we examine some generic features of the moduli dynamics in the
early universe to see how it
depends on the parameters involved and also
identify the initial moduli misalignment
$\delta\phi$.
We then use the known cosmological and astrophysical observations to obtain an
upper bound of the moduli misalignment $\delta\phi$ in section IV.
In this regard, we first consider the case that there is no entropy production
after the moduli oscillation begins and summarize the results in Figure 1.
As is well known, the light moduli density (relative to the entropy density)
can be diluted if there occurs an entropy production after the moduli
oscillation begins.  We thus finally discuss the dilution of the light moduli
density in  various possible cosmological scenarios with a late entropy
production, including the case that the entropy-producing field $\varphi$ is
a massive moduli with $m_{\varphi}\gtrsim {\cal O}(40)$ TeV and also the case
that $\varphi$ corresponds to the flaton field triggering thermal inflation
\cite{lyth}.
The relaxed bounds on $\delta\phi$ for the cases with a late entropy
production are depicted in Figures 2 and 3.

\section{Light axion-like moduli in string/M-theory}

To make our motivation more clear, let us discuss in more detail the masses
of axion-like moduli in compactified  string/$M$-theory.
As is well known, the theory predicts  the dilaton superfield $S$ and the
K\"ahler moduli superfields ${\cal T}_I$ ($I=1,2,...,h_{1,1}$) for generic
compactifications preserving the four-dimensional supersymmetry \cite{gswbook}.
The scalar components ${\rm Re}(S)$ and ${\rm Re}({\cal T}_I)$
of these superfields determine the four-dimensional gauge coupling
constant and the size (and also the shape) of the internal
six manifold, respectively.
For an isotropic six  manifold, we have
$\langle {\rm Re}({\cal T}_I)\rangle
\approx \langle {\rm Re}({\cal T})\rangle$ where 
${\rm Re}({\cal T})$ denotes the overall modulus
whose VEV  corresponds to the
radius-squared of the internal six manifold in the heterotic string
length  unit.
The pseudoscalar components ${\rm Im}(S)$ and ${\rm Im}({\cal T}_I)$ 
are often called 
the model-independent axion and the model-dependent K\"ahler axions,
respectively \cite{witten,choi1}.
These axion-like moduli are periodic variables and we normalize them by
imposing the periodicity conditions:
\begin{equation}
{\rm Im}(S)\equiv {\rm Im}(S)+1,
\quad
{\rm Im}({\cal T}_I)\equiv {\rm Im}({\cal T}_I)+1.
\label{eq:periodicity-condition}
\end{equation}
Under this normalization, the scalar components are given by \cite{choi} 
\begin{eqnarray}
{\rm Re}(S)&\approx & {1\over\alpha_{GUT}}
\approx 
4\pi e^{-2 D}\frac{V}{(2\alpha^{\prime})^3}
\approx 2 (4\pi\kappa^2)^{-2/3}V\ ,
\nonumber \\
{\rm Re}({\cal T})&\approx &
\frac{6^{1/3}}{4\pi^2}\frac{V^{1/3}}{2\alpha^{\prime}}
\approx 6^{1/3}(4\pi
\kappa^2)^{-1/3}\pi \rho V^{1/3}\ ,
\label{dilaton-modulus}
\end{eqnarray}
where $e^{2D}$ is the heterotic string dilaton,
$V$ is the internal space volume, 
$\kappa^2$ is the eleven-dimensional gravitational
coupling constant, and finally
$\pi \rho$ denotes the length of the eleventh
segment in $M$-theory. 
The above relations, together with $M_P=2\pi\rho \kappa^{-2} V$, 
show that the  heterotic string coupling $e^{2 D}$ scales
as $\alpha_{GUT}[{\rm Re}({\cal T})]^3$  and the eleventh length
$\rho$ scales as $M_P^{-1}[{\rm Re}({\cal T})]^{3/2}$.
Inserting the proper numerical coefficients,
it is easy to see that for $\alpha_{GUT}\approx 1/25$ and $M_P\approx
2\times 10^{18}$ GeV,
the large radius limit with $\langle{\rm Re}({\cal T})\rangle\gg1$ 
corresponds to  
the $M$-theory limit 
with a strong heterotic string coupling,
which can be 
described by an eleven-dimensional supergravity on
a manifold with boundary
\cite{horava}.

For the superfield normalization determined by   
the periodicity condition (\ref{eq:periodicity-condition}), 
the holomorphic gauge kinetic functions satisfy the relation \cite{choi}
\begin{equation}
4\pi f_{E_8}-4\pi f_{E^{\prime}_8}\approx
\sum_I l_I {\cal T}_I,
\label{eq:relation-2}
\end{equation}
where $l_I$'s are integer coefficients, and $f_{E_8}$ and $f_{E^{\prime}_8}$
denote the gauge kinetic functions for $E_8$ and $E^{\prime}_8$, respectively.
Here the gauge kinetic functions are normalized as
$\langle {\rm Re}(f_a)\rangle =1/g_a^2$ and
$\langle {\rm Im}(f_a)\rangle =\theta_a/8\pi^2$,
where $g_a$ and  $\theta_a$ denote the gauge coupling constant and 
the vacuum angle for the $a$-th gauge group, respectively.
Note that  for integer $l_I$ the relation (\ref{eq:relation-2}) is 
consistent with 
the periodicity condition (\ref{eq:periodicity-condition}) 
and the periodic vacuum angles $\theta_a\equiv\theta_a+2\pi$.   
The  relation (\ref{eq:relation-2}) suggests that,
even in the $M$-theory limit, ${\rm Re}({\cal T})\approx {\rm Re}({\cal T}_I)$ 
can {\it not} be
arbitrarily large, but is constrained {\it not} to significantly exceed
$4\pi{\rm Re}(f_{E_8})\approx {1\over \alpha_{GUT}}$.
For compactifications on a smooth Calabi-Yau
manifold with vanishing $E^{\prime}_8$ field-strength,
it turns out that at least
one of $l_I$'s is positive and other $l_I$'s are still non-negative
integers \cite{choi}. 
This then leads to the upper limit:
\begin{equation}
\langle {\rm Re}({\cal T})\rangle\lesssim {1\over \alpha_{GUT}},
\label{eq:tbound}
\end{equation}
which  corresponds to the lower
limit on the Newton's  constant $G_N$ discussed in \cite{witten1}, 
and also to the
lower limit on the Kaluza-Klein scale $M_{KK}\gtrsim\alpha_{GUT}/\sqrt{G_N}$
discussed in \cite{kaplu}.

The K\"ahler potential of the effective supergravity model depends upon
${\rm Re}(S)$ and also ${\rm Re}({\cal T}_I)$ with unsuppressed coefficients 
of order unity. As a result,  once the four-dimensional
supersymmetry is broken, ${\rm Re}(S)$ and ${\rm Re}({\cal T}_I)$ 
receive the masses of order $m_{3/2}$ from the supergravity scalar potential.
However the masses of the axion-like moduli ${\rm Im}(S)$ and 
${\rm Im}({\cal T}_I)$
are constrained by the approximate non-linear global $U(1)$ symmetries
defined as
\begin{equation}
U(1)_S: \, {\rm Im}(S)\rightarrow {\rm Im}(S)+\alpha_S,
\quad
U(1)_{I}:\, {\rm Im}({\cal T}_I)\rightarrow {\rm Im}({\cal T}_I)+\alpha_I,
\end{equation}
where $\alpha_S$ and $\alpha_I$ denote arbitrary real constants.
In the limit where one (combination) of these $U(1)$'s becomes an exact
symmetry, the corresponding axion-like field becomes an exact Goldstone
boson and thus
is massless.  In string/$M$-theory, these $U(1)$-symmetries are explicitly
broken either by the Yang-Mills axial anomaly or by the world-sheet
(membrane) instanton effects \cite{dine}.
If the hidden sector gauge interactions provide a dynamical seed for
supersymmetry breaking, the $U(1)$-breaking by the hidden sector Yang-Mills
axial anomaly is so strong that the linear combination of ${\rm Im}(S)$ and
${\rm Im}({\cal T}_I)$ which couples to the hidden sector anomaly get the masses of
order $m_{3/2}$.  However most of the known compactification models allow a
combination of $U(1)_S$ and $U(1)_{I}$ which is {\it free from} the hidden
sector anomaly, but still explicitly broken by the world-sheet instanton
effects and/or by the observable sector anomaly (mainly the QCD anomaly)
\cite{choi1,dine}.  

Let $\phi$, being a linear combination of ${\rm Im}(S)$ and 
${\rm Im}({\cal T}_I)$,
denote the axion-like moduli for the combination
of $U(1)_S$ and $U(1)_{I}$ which is free from the hidden
sector anomaly.   Then the  effective potential $V_{\phi}$ of
this axion-like moduli includes  first of all the contribution 
from the world-sheet instanton
effects which is estimated to be \cite{bank,choi}
\begin{equation}
V_{\rm WS}\approx e^{-2\pi \langle {\rm Re}({\cal T})\rangle}m_{3/2}^2M_P^2.
\label{eq:world}
\end{equation}
If $\phi$ couples to the observable sector QCD anomaly, 
$V_{\phi}$ would include also the contribution  from the QCD anomaly,
$V_{QCD}\approx f_{\pi}^2m_{\pi}^2$.
In fact, one could argue based on supersymmetry and the periodicity
condition (4) for the axion-like moduli that there is no other type of
contribution to $V_{\phi}$ \cite{choi}.

For  compactifications with 
$\langle{\rm Re}({\cal T})\rangle\lesssim17$,
$V_{\phi}$ is dominated by the world-sheet instanton contribution
(\ref{eq:world}).
(Here and in the following, we assume  $m_{3/2}\approx 1$ TeV for the
simplicity of the discussion.)
Then the mass of the axion-like moduli field $\phi$ is estimated to be
\begin{equation}
m_{\phi}\approx e^{-\pi\langle {\rm Re}({\cal T})\rangle} m_{3/2}.
\label{eq:ws-mass}
\end{equation}
However if $\langle{\rm Re}({\cal T})\rangle\gtrsim17$ and also $\phi$ couples
to the QCD anomaly, $V_{\phi}$ is dominated by the QCD contribution, leading
to $m_{\phi}\approx f_{\pi}m_{\pi}/M_P$.
(If $\langle {\rm Re}({\cal T})\rangle$ is even bigger than about 20,
we have $V_{WS}\lesssim 10^{-9} V_{QCD}$ and then this $\phi$ can be
identified as the QCD axion solving the strong CP problem \cite{bank,choi}.)
In models with $h_{1,1}>1$, there can be a combination of $U(1)_S$ and
$U(1)_I$ which is free from {\it both} the hidden sector Yang-Mills
axial anomaly and the observable sector QCD anomaly.
The axion-like moduli field for this combination does {\it not} couple
to the QCD anomaly and then its mass is given by Eq.~(\ref{eq:ws-mass}) 
even for $\langle{\rm Re}({\cal T})\rangle\gtrsim 17$.

The above discussion, particularly Eq.~(\ref{eq:ws-mass}), implies that
$m_{\phi}$ is highly sensitive to the compactification radius which is
measured by $\langle{\rm Re}({\cal T})\rangle$. 
It can be extremely smaller than $m_{3/2}$ if the compactification radius
is large enough to have $\langle {\rm Re}({\cal T})\rangle\gg 1$.
As was noted in the discussions above (\ref{eq:tbound}), 
a large value of $\langle{\rm Re}({\cal T})\rangle$ 
is allowed
in the $M$-theory limit but it is constrained {\it not} to significantly
exceed ${1\over \alpha_{GUT}}\approx 25$.  Then for the range
$0<\langle{\rm Re}({\cal T})\rangle\lesssim 25$, 
$m_{\phi}$ can be anywhere between
$m_{3/2}$ and the extremely small mass $10^{-34} m_{3/2}$. 
Note that even when $\langle{\rm Re}({\cal T})\rangle\approx 1$ 
for which the
weakly-coupled heterotic string theory provides a sensible description
for the dynamics above $M_P$, the axion-like moduli mass $m_{\phi}$ can be
smaller than $m_{3/2}$ by one or two orders of magnitudes.

\section{Moduli dynamics in the early universe}

In the previous section, we have noted that, in string/$M$-theory 
with a large
compactification radius, some of the axion-like moduli can be much lighter
than the gravitino mass.  Motivated by this observation,
in this and next sections,  we study cosmological aspects of a
generic light modulus $\phi$ with an arbitrary mass below the weak scale.

Moduli dynamics in the early universe would be governed by the free energy
density $V_{\rm eff}$ which  depends not only on
$\phi$ but also on other
scalar fields $\Phi$ and the radiation temperature $T$.  
Expanding
$V_{\rm eff}$ around the {\it present} moduli VEV which is set to zero,
one generically has  
\begin{equation}
V_{\rm eff}(\phi,\Phi,T)= \Omega_0(\Phi,T)
 + \Omega_1(\Phi,T)\phi
 + \frac12\Omega_2(\Phi,T)\phi^2
 + ...,
\label{free-energy}
\end{equation}  
where the moduli-tadpole $\Omega_1$ arises since the expansion is made 
around the present moduli VEV, {\it not} around the minimum of
$V_{\rm eff}$ in the early universe.
Obviously at present, $\Omega_n$ are given by
\begin{equation}
\Omega_0(\Phi_0, T_0) = \Omega_1(\Phi_0, T_0 )= 0 \ ,
\quad
\Omega_2(\Phi_0, T_0 )=m_{\phi}^2 \ ,
\end{equation}
where $\Phi_0$, $T_0$, and $m_{\phi}$ denote the {\it present} VEV of $\Phi$,
the {\it present} temperature, and the {\it present} mass of $\phi$, 
respectively, and  the vanishing of $\Omega_0(\Phi_0,T_0)$ comes from
the vanishing (or negligibly small) cosmological constant at present. 

In the early universe,  $\Phi$ and  $T$ can take
values far away from the present ones. As a result,
$\Omega_n$'s in the early universe can
significantly differ from their present values. 
For instance, when $H\gg m_{\phi}$, the effective moduli mass
$\sqrt{{\Omega_2}}$ can be of order $H$
and thus much bigger than $m_{\phi}$.

To proceed, let us parameterize the  free energy density
(\ref{free-energy})
as follows \cite{lyst}:
\begin{equation}
V_{\rm eff}(\phi,T) = \frac{1}{2}m_\phi^2\phi^2
                     +\frac{1}{2}c^2H^2(\phi-\phi_1)^2 +...\ ,
\label{eq:effective-potential}
\end{equation}
where $m_{\phi}$ denotes the moduli mass {\it at present} 
and the ellipsis
denote the irrelevant terms.
This parameterization is useful since
in most cases of interest, $c$ and $\phi_1$ are approximately  
time-independent constants, which would greatly simplify the analysis.
Obviously the coefficients $c$
and $\phi_1$  describes the 
non-derivative interactions of $\phi$ with the background 
energy density
in the early universe,
e.g. the radiation or the inflaton energy density.
Generically $\phi_1$ is expected to be of order $M_P$.
The value of $c$ measures how strongly $\phi$ couples
to supersymmetry breaking environment  and thus
\begin{equation}
c\approx m_{\phi}/m_{3/2}.
\end{equation}
It was noted in Ref.~\cite{linde} that in the case of $c\gg1$, the initial
moduli misalignment  can be rapidly damped away.
In this paper, we consider only the case
$c\lesssim {\cal O}(1)$ which appears to be more natural.
Note that for axion-like moduli discussed in the previous section,
$c\approx e^{-\pi {\rm Re}({\cal T})}$ 
can be very small.

Treating $c$ and $\phi_1$ as constants,
the equation of motion for $\phi$ is given by
\begin{equation}
\ddot\phi + 3H\dot\phi + (m_\phi^2 + c^2H^2)\phi = c^2H^2\phi_1 \ .
\label{eq:equation-of-motion}
\end{equation}
In the following, we
wish to examine the evolution of $\phi$ by solving this equation with the
initial value taken at some initial time $t_i$:
\begin{equation}
\phi(t_i)=\phi_i \ , \hspace{5mm}
\dot\phi(t_i)=0 \ .
\label{eq:initial-condition}
\end{equation}

Let us first consider the evolution of $\phi$ 
during inflationary era for which the Hubble parameter $H$ 
is roughly a constant.
The solution of Eqs.~(\ref{eq:equation-of-motion}) and
(\ref{eq:initial-condition}) is easily found to be as
follows:
For $(\frac32H)^2>m_\phi^2+c^2H^2$,
\begin{equation}
\phi(t) = \phi_{\rm min} + (\phi_i-\phi_{\rm min})
\left[ \frac{1+\beta}{2\beta}e^{-\frac{3(1-\beta)}{2}H(t-t_0)}
      -\frac{1-\beta}{2\beta}e^{-\frac{3(1+\beta)}{2}H(t-t_0)} \right],
\end{equation}
while for $(\frac32H)^2<m_\phi^2+c^2H^2$,
\begin{equation}
\phi(t) = \phi_{\rm min} + (\phi_i-\phi_{\rm min})
e^{-\frac32H(t-t_0)}\left[ \cos [\beta' m_\phi(t-t_0)]
-\frac{3H}{2\beta' m_\phi}\sin [\beta' m_\phi(t-t_0)] \right] \ ,
\label{solution-3}
\end{equation}
where $\beta=\sqrt{1-\frac49(c^2+\frac{m_\phi^2}{H^2})}$,
$\beta'=\sqrt{1-(\frac94-c^2)\frac{H^2}{m_\phi^2}}$,
and $\phi_{\rm min}$ is the temporal minimum of the effective
potential which is given by
\begin{equation}  \label{phimin}
\phi_{\rm min} = \frac{c^2H^2}{m_\phi^2+c^2H^2}\phi_1 \ .
\end{equation}

Let us consider some 
interesting limits of the above solution.
If $m_\phi\ll H$ and $c\ll 1$, we have $\beta\approx 1-\frac{2}{9}(c^2
+\frac{m_{\phi}^2}{H^2})$ and then the modulus value after the inflation
($\phi_f$) is given by
\begin{equation}
\phi_f\approx \phi_i-\frac{N_e}{3}(\phi_i-\phi_{\rm min})\left( H^2\over
m_{\phi}^2+c^2H^2\right),
\end{equation}
where $N_e$ is the number of e-folding.
In the above, we assumed that $c$ and $m_{\phi}/H$ are small enough
so that $N_e\ll {H^2\over c^2H^2+m_{\phi}^2}$.
Thus simply speaking, in the case of $m_\phi\ll H$ and $c \ll 1$,
$\phi$ is frozen at its initial value $\phi_i$.
This is what we thought to happen during inflation before realizing the role
of $H^2$ term in Eq.~(\ref{eq:effective-potential}).
But for $m_\phi\ll H$ and $c\sim 1$,
we have the final modulus value
\begin{equation}
\phi_f\approx \phi_{\rm min}+(\phi_i-\phi_{\rm min})\times
{\cal O}(e^{-3N_e/2}),
\label{eq:18}
\end{equation}
showing that $\phi$ rapidly approaches to
the temporal minimum $\phi_{\rm min}\simeq \phi_1$
{\it independently of} the value $\phi_i$ before the inflation.  
Finally for $m_{\phi}\gg H$, $\phi$ exponentially approaches
to $\phi_{\rm min}\simeq\frac{c^2H^2}{m_\phi^2}\phi_1$,
but with an exponentially decreasing oscillatory tail.
This is what happens in so-called thermal inflation \cite{lyth}
which will be discussed in more detail in the next section.

Let us now consider the evolution of $\phi$ 
during the radiation-dominated (RD) or matter-dominated (MD) era
for which
the Hubble parameter is given by
$H=p/t$ where $p=\frac12$ (RD), $\frac23$ (MD).
In RD or MD era,
the solution of Eq.~(\ref{eq:equation-of-motion})
is given by
\begin{equation}
\phi(z) = p^2c^2\phi_1\frac{S_{\alpha-1,\nu}(z)}{z^\alpha} +
          C_1\frac{J_\nu(z)}{z^\alpha} + C_2\frac{Y_\nu(z)}{z^\alpha} \ ,
\label{eq:solution-phi}
\end{equation}
where  $\nu^2=\alpha^2-p^2c^2\ge0$ for $\alpha=\frac12(3p-1)$
and  $z=m_\phi t$.
Here $J_\nu(z)$ and $Y_\nu(z)$ are Bessel functions, and
$S_{\mu,\nu}(z)$ is Lommel function which is defined by
\begin{eqnarray}
S_{\mu,\nu}(z)&& = \frac{\pi}{2}\left[
     Y_\nu(z)\int_0^z y^\mu J_\nu(y)\,dy
    -J_\nu(z)\int_0^z y^\mu Y_\nu(y)\,dy \right]
\nonumber\\ &&
+2^{\mu-1} \Gamma(\frac{\mu-\nu+1}{2})\Gamma(\frac{\mu+\nu+1}{2})\left\{
\sin(\frac{\mu-\nu}{2}\pi)J_\nu(z)-\cos(\frac{\mu-\nu}{2}\pi)Y_\nu(z) \right\}
\ .
\end{eqnarray}
The coefficients $C_1$ and  $C_2$ are fixed by the initial condition
(\ref{eq:initial-condition}):
\begin{eqnarray} \label{Cs}
C_1 &=& 
A_1 \phi_i + B_1 c^2\phi_1 \ , \nonumber \\
C_2 &=& 
A_2 \phi_i + B_2 c^2\phi_1 \, 
\label{eq:integral-constants}
\end{eqnarray}
where
\begin{eqnarray}
A_1 &=& 
\frac{\pi}{2}z_i^\alpha \left\{z_iY'_\nu(z_i)-\alpha Y_\nu(z_i)\right\}
\nonumber \\
A_2 &=&
-\frac{\pi}{2}z_i^\alpha \left\{z_iJ'_\nu(z_i)-\alpha J_\nu(z_i)\right\}
\nonumber \\
B_1 &=&
-\frac{\pi}{2}z_i \left\{Y'_\nu(z_i)S_{\alpha-1,\nu}(z_i)
             -Y_\nu(z_i)S'_{\alpha-1,\nu}(z_i)\right\}p^2
\nonumber \\
B_2 &=&
\frac{\pi}{2}z_i \left\{J'_\nu(z_i)S_{\alpha-1,\nu}(z_i)
             -J_\nu(z_i)S'_{\alpha-1,\nu}(z_i)\right\}p^2.
\label{eq:coefficients}
\end{eqnarray}
Here $z_i=m_{\phi} t_i$ 
and the prime denotes the differentiation with respect to $z$.

From the solution (\ref{eq:solution-phi}),
we can calculate the moduli abundance coming from
the coherent oscillation.
For $z\gg1$, $\phi(z)$ is dominated  by the oscillating tail
\begin{equation}
\phi(z) \approx \left(\frac{2}{\pi}\right)^{1/2} z^{-\frac32p}
\left\{
C_1 \cos[z-(\nu+\frac12)\frac{\pi}{2}] +
C_2 \sin[z-(\nu+\frac12)\frac{\pi}{2}]
\right\} \ .
\label{eq:moduli-oscillation}
\end{equation}
The energy density of this oscillating modulus  is given by
\begin{equation}
\rho_\phi=\frac12m_\phi^2(\phi'^2+\phi^2)\approx
\left(\frac{2}{\pi}\right)(C_1^2+C_2^2)m_\phi^2z^{-3p}
\label{eq:moduli-energy}
\end{equation}
and normalizing it  by the entropy density
$s=\frac{2\pi}{45}g_*T^3$, we  find
\begin{equation}
m_\phi Y_\phi \equiv {\rho_{\phi}\over s}= 
\frac{45}{2\pi^2g_*} (C_1^2+C_2^2) \frac{m_\phi^2}{z^{3p}T^3} \ ,
\label{mY}
\end{equation}
 where $g_*$ denotes the effective number of
the relativistic degrees of freedom at $T$.

Most of the cosmological implications of $\phi$ is in fact determined
by the oscillation amplitude:
\begin{equation}
\delta\phi\equiv\left(\frac{C_1^2+C_2^2}{\pi}\right)^{1/2}
\label{eq:def-misalignment}
\end{equation}
which we  call the initial moduli misalignment throughout this paper.
The oscillation coefficients $C_1$ and $C_2$ are determined by
the  modulus value $\phi_i=\phi(t_i)$ at an initial time $t_i$ 
with $\phi^{\prime}(t_i)=0$,
and also by the two dynamical parameters
$c$ and $\phi_1$ in
the free energy density (\ref{eq:effective-potential}) which governs
the moduli dynamics at later time $t\gtrsim t_i$.
Depending upon the periods under consideration,
$z_i=m_{\phi}t_i$ may be chosen to be either
very small, or of order unity, or very large.
It turns out that the coefficients $A_{1,2}$ and $B_{1,2}$ 
in Eq.~(\ref{eq:coefficients}) are essentially of order one for
$z_i\sim 1$ and an arbitrary value of $c\lesssim 1$.
Thus roughly speaking, $C_1$ and $C_2$ are
linear combinations
of $\phi_i$ (at $t_i\approx m_{\phi}^{-1}$)
and $c^2\phi_1$ with coefficients of order one.
At any rate, using Eqs.~(\ref{eq:def-misalignment}) and
(\ref{eq:integral-constants}),
any cosmological bound on $\delta\phi$
can be translated into a constraint 
on the parameter set $(\phi_i, c^2\phi_1)$.

If the universe were radiation-dominated when $\phi$-oscillation
begins at $t\sim m_{\phi}^{-1}$ and there is no entropy production
since then,
we have $z=\left(\frac{45}{2\pi^2g_*}\right)^{1/2}\frac{m_\phi M_P}{T^2}$,
and thus 
\begin{equation}
m_\phi Y_\phi = \left(\frac{45}{2\pi^2g_*}\right)^{1/4}
\left(\frac{\delta\phi}{M_P}\right)^2 (m_\phi M_P)^{1/2} \ .
\label{mY-RD}
\end{equation}
In other case that  the universe was matter-dominated,
for instance by the inflaton oscillation or by the heavy moduli oscillation,
at the moment when $\phi$-oscillation begins,
one has to take into account the subsequent entropy production
due to the out-of-equilibrium decays of the inflaton or heavy moduli.
Assuming that the whole matter energy
is converted  into the radiation 
with the reheat temperature $T_R$,
the modulus energy density normalized by the entropy density is given by
Eq.~(\ref{mY}) evaluated at the reheat time 
$t_R
=\left(\frac{40}{\pi^2g_*}\right)^{1/2}\frac{M_P}{T_R^2}$,
\begin{equation}
m_\phi Y_\phi = \frac{3}{4} \left(\frac{\delta\phi}{M_P}\right)^2 T_R \ .
\label{mY-MD}
\end{equation}

\section{The constraints on the initial misalignment}

\subsection{The cosmological moduli problem}

In the previous section, we obtained the moduli energy density normalized by
the entropy density when the coherent oscillation begins during the RD era:
\begin{equation}
m_{\phi} Y_{\phi}\approx g_*^{-1/4}\left({\delta\phi\over M_P}\right)^2
(m_{\phi}M_P)^{1/2}
\approx 6\times 10^8 \left(\frac{m_\phi}{{\rm GeV}}\right)^{1/2}
\left(\frac{\delta\phi}{M_P}\right)^2 \, {\rm GeV} \ .
\end{equation}
A coherently oscillating modulus can dominate the energy density of the
universe {\it unless}  $m_{\phi}Y_{\phi}$ is less than the 
temperature $T_{EQ}\approx3\eV$  of matter-radiation equality.
This implies that one has to worry about
the over-produced moduli which would contradict with the cosmological
observations {\it as long as} the moduli mass is in the range
\begin{equation}
m_\phi \gtrsim 10^{-26} \eV \, . 
\label{aslongas}
\end{equation}
On the other hand, sufficiently heavy moduli decaying before about one second
do not affect the standard prediction of the big-bang nucleosynthesis, and
thus would not contradict with the currently known cosmological observations.
Having interactions suppressed by $M_P$, the moduli lifetime is estimated to be
\begin{equation}
\tau_\phi \approx
\xi\times10^{14}\left(\frac{m_\phi}{\GeV}\right)^{-3} {\rm sec} \ ,
\label{lifetime}
\end{equation}
where $\xi$, being roughly of order one, is a coefficient which accounts for
the ambiguity  in our estimate of the lifetime.
For moduli lighter than $\sim40\TeV$, their decay products may change
the abundance of the light elements after or during the nucleosynthesis,
the spectrum of cosmic background radiation, or  
the observed $\gamma$ and $X$-ray backgrounds \cite{ellis}.
Moduli whose lifetime is longer than the age of the universe
would overclose the universe {\it unless} $m_\phi Y_\phi\lesssim3\eV$.
This means for the moduli mass $m_\phi\lesssim0.1\GeV$, the initial
misalignment is constrained as 
\begin{equation}
\frac{\delta\phi}{M_P} \lesssim
4\times10^{-9}\left(\frac{m_\phi}{0.1\GeV}\right)^{-1/4} \ .
\label{critical}
\end{equation}
{\it unless} there is a late entropy production after $\phi$ starts to
oscillate.

Based on the analysis of \cite{ellis}, the recently reported $X$-ray
background \cite{Xray} and Eq.~(\ref{critical}),
we obtain the constraints on $\delta\phi$ arising from these considerations
again under the assumption that there is {\it no} entropy production after
$\phi$-oscillation begins at $t\sim m_{\phi}^{-1}$.
The results are summarized in Figure 1 showing the cosmological upper
limit on the initial moduli misalignment $\delta\phi$ for the moduli mass
$m_\phi$ below $\sim40\TeV$.
The line (0) comes from Eq.~(\ref{critical}) which is required for the moduli
{\it not} to overclose the universe,
and the line (1) from the recently reported $X$-ray background.
The lines (a)-(h) are obtained from the observed  $\gamma$-ray background,
the spectrum of cosmic microwave background radiation,
and also the light element abundances.

The results of Figure 1 show that, in the absence of a late time entropy
production, the initial moduli misalignment is required to be very small
compared to its natural value $\sim M_P$, for instance
$\delta \phi\lesssim 10^{-6}  M_P$ for $m_{\phi}\approx 1$ eV,
$\delta \phi\lesssim 10^{-10} M_P$ for $m_{\phi}\approx 1$ MeV,
$\delta \phi\lesssim 10^{-11} M_P$ for $m_{\phi}\approx 1$ GeV.
Let us recall that $\delta\phi$ is determined by
the modulus value $\phi_i$ at an initial time $t_i$
with $\phi^{\prime}(t_i)=0$,
and also the dynamical parameters $(c,\phi_1)$
in the free energy density which would govern the moduli dynamics
at time $t\gtrsim t_i$.
(See Eqs.~(\ref{eq:integral-constants})--(\ref{eq:def-misalignment}).)
Its order of magnitude is roughly given
by the bigger one among 
$\phi_i$ and $c^2\phi_1$ for the initial time $t_i\approx m_{\phi}^{-1}$. 
Its natural value would be of order $M_P$ which then falls at far
above the upper limits in Figure 1 for most range of the moduli mass.

\subsection{Dilution by heavy moduli decays}

If there is an entropy production during the period after $\phi$ starts
to oscillate but well before $\phi$ decays, the moduli energy density is 
diluted as $Y_\phi \to Y_\phi/\Delta$ where
$\Delta=S_{\rm after}/S_{\rm before}$ denotes the entropy production factor.
Since $Y_{\phi}\propto \delta\phi^2$, this obviously leads to the relaxation
of the constraints on the initial moduli misalignment which will be
discussed below.

Usually the entropy production is due to out-of-equilibrium decays of
non-relativistic particles which appear in the form of another coherently
oscillating scalar field $\varphi$. 
In order to be compatible with the big-bang nucleosynthesis, this
entropy-producing scalar field $\varphi$ is required to decay before the
nucleosynthesis with the reheat temperature $T_R\gtrsim6\MeV$ \cite{laza}.
Also note that if the light moduli $\phi$ is so light that its oscillation
begins after the entropy production by $\varphi$ is over,
i.e. $m_{\phi}\lesssim g_*^{1/2}T_R^2/M_P$,
$Y_{\phi}$ is not affected by the entropy production by $\varphi$.
Thus the energy density of oscillating $\phi$ can be diluted
{\it only} for the moduli mass
\begin{equation}
m_{\phi}\gtrsim 5\times10^{-14}\left(\frac{T_R}{6\MeV}\right)^2\eV \ .
\label{hmd-bound}
\end{equation}

To be more specific, let us consider the interesting possibility that
the entropy-producing field $\varphi$ is a massive moduli $\varphi$
with $\delta\varphi\sim M_P$ and $m_\varphi\sim 40\TeV$ which would give
a maximal entropy production with $T_R\approx6\MeV$.
This heavy modulus dominates the energy density of the universe as soon as
its coherent oscillation begins at $T\approx10^{11}\GeV$.
The light moduli which are lighter than $\sim40\TeV$ but heavier than the
bound in Eq.~(\ref{hmd-bound}) start to coherently oscillate during the matter
dominated era by the heavy modulus oscillation.
Then we can apply Eq.~(\ref{mY-MD}) and get the abundance
\begin{equation}
m_\phi Y_\phi \approx T_R\left(\frac{\delta\phi}{M_P}\right)^2 \ ,
\label{mY_hmd}
\end{equation}
which is diluted compared to Eq.~(\ref{mY-RD}) by a factor
$S_{\rm after}/S_{\rm before}=(m_\phi M_P)^{1/2}/g_*^{1/4}T_R$.
It is rather straightforward to derive the cosmological limits
on the light moduli misalignment $\delta\phi$
for the diluted moduli density (\ref{mY_hmd})
as we did for the case without any late entropy production.
The results for 
the case of $T_R=6\MeV$, i.e. maximal dilution, are shown in Figure~2.
It shows that the entropy production by heavy moduli decays ameliorates
but does not completely solve the cosmological moduli problem.

\subsection{Dilution by thermal inflation}

The most efficient way to dilute the dangerous light moduli $\phi$ is to have
a late inflation since inflation dilutes the moduli density both by the
spatial expansion and by the large amount entropy production.
The most natural framework for a late inflation would be the so-called thermal
inflation models \cite{lyth}.

In thermal inflation models, the entropy-producing field $\varphi$ 
corresponds to a flaton field parameterizing  a flat direction in
supersymmetric models.  This flat direction  is lifted by the soft
breaking mass and also by the Planck scale suppressed
non-renormalizable terms, leading to the following (renormalization
group improved) effective potential at zero temperature:
\begin{equation}
V_{\varphi}=V_0-m_{\varphi}^2|\varphi|^2+\frac{|\varphi|^{2n+4}}{M_P^{2n}},
\end{equation}
where $n$ is a model-dependent integer and the negative mass-squared can
arise as a consequence of radiative corrections associated with 
the strong Yukawa coupling of $\varphi$.
The true vacuum expectation value is
$\langle\varphi\rangle\approx(m_\varphi M_P^n)^{1/(n+1)}$
and $V_0$ is adjusted to
$V_0\approx m_\varphi^2\langle\varphi\rangle^2
    \approx m_\varphi^{\frac{2n+4}{n+1}}M_P^{\frac{2n}{n+1}}$
in order for the true vacuum energy density to vanish.
At high temperature $T\gg m_\varphi$, the effective flaton mass-squared 
including the thermal contribution of ${\cal O}(T^2)$ is positive and thus
$\langle\varphi\rangle_{T\gg m_{\varphi}}=0$.
For the period of $m_\varphi\lesssim T\lesssim V_0^{1/4}$, the universe
is vacuum-dominated, yielding an exponential expansion with  
the number of e-foldings ($N_e$) which is determined by  
$e^{N_e}\approx0.42(100/g_*)^{1/4}(M_P/m_\varphi)^{n/2n+2}$.
At $T\approx m_\varphi$, the flaton starts to roll down to its true vacuum
value $\langle\varphi\rangle$, and then the vacuum energy density $V_0$ is 
converted into the energy density of coherently oscillating flaton field.
The oscillating flatons eventually decay and are converted into the radiation
with the reheat temperature
\begin{equation}
T_R \approx 1.7g_R^{-1/4}\sqrt{M_P\Gamma_\varphi}
    \approx 0.1\gamma^{1/2}g_R^{-1/4}m_\varphi
            \left(\frac{m_\varphi}{M_P}\right)^{\frac{n-1}{2n+2}},
\label{reheating-temperature}
\end{equation}
where $\gamma$ is introduced to parameterize the $\varphi$-decay width
$\Gamma_{\varphi}=\gamma m_{\varphi}^3/64\pi\langle\varphi\rangle^2$
and $g_R$ denotes the effective number of the relativistic degrees of
freedom at $T_R$.
As was discussed in \cite{chun}, $\varphi$ can couple to ordinary matter
through its mixing with Higgses, and then the most efficient decay channel
is the decay into stop and anti-stop pair.
Assuming  $m_\varphi> 2m_{\tilde{t}}$ so that $\varphi$ can decay into stop
pairs $\tilde{t}$ and $\tilde{t}^*$, we have roughly
$\gamma\approx(2m_{\tilde{t}}/m_{\varphi})^4$.
Although it can be a quite small number (particularly when
$m_{\varphi}\gg m_{\tilde{t}}$), 
we assume here $\gamma\sim 1$ as a conservative choice.  

The entropy production factor of thermal inflation is given by
\begin{equation}
\frac{S_{\rm after}}{S_{\rm before}}
\approx \frac{V_0}{3T_Rm_\varphi^3}
       \approx 0.1\gamma^{-\frac12}
               \left(\frac{M_P}{m_\varphi}\right)^{\frac{5n-1}{2n+2}}
\ .
\label{entropy-production-factor}
\end{equation}
{}From Eq.~(\ref{reheating-temperature}) we see that
to achieve $T_R\gtrsim6\MeV$ it is required that
$\gamma^{1/2}m_\varphi\gtrsim100\MeV$,
$\gamma^{3/7}m_\varphi\gtrsim 60\GeV$, and
$\gamma^{2/5}m_\varphi\gtrsim700\GeV$ for $n=1,2,3$ respectively.
Under this restriction, the maximal entropy production factor is
$2\times10^{18}$,
$6\times10^{23}\gamma^{1/7}$, 
$8\times10^{27}\gamma^{1/5}$
respectively.

The light moduli $\phi$ can start to oscillate either before or after the
thermal inflation depending upon their masses.
Obviously the moduli oscillation should start before thermal inflation occurs
in order for thermal inflation to sufficiently dilute the moduli density.
This gives a lower bound on the moduli mass for which thermal inflation leads
to the sufficient dilution of moduli density:
$m_\phi\gtrsim H_{\rm TI} \approx (V_0/M_P^2)^{1/2}
                          \approx 100\eV,\ 100\keV,\ 1\MeV$
for $n=1,2,3$ respectively, where $H_{\rm TI}$ is the Hubble
expansion parameter during thermal inflation.
The moduli lighter than this bound but start to oscillate before the flaton
decay are also diluted somewhat by the entropy production due to the flaton
decay.  For this mass range the analysis of the previous section can be
applied.  For $m_\phi\approx H_{\rm TI}$, we need a detailed analysis,
but we will not concern such details.
In Figure~3(a), we showed the relaxed constraints on the initial misalignment
$\delta\phi$ when the moduli 
density is maximally diluted by thermal inflation.

Although thermal inflation provides a huge entropy and thus dilute
the moduli energy density due to the intial  misalignment,
it can cause an additional misalignment induced by a shifted minimum
of the free energy density during the inflation.
As was noted in the discussions below Eq.~(\ref{eq:18}),
during the thermal inflation period, the shifted minimum is given by
$\phi_{\rm min}\approx\frac{c^2H_{\rm TI}^2}{m_\phi^2}\phi_1$.
For $H_{\rm TI}\lesssim m_{\phi}$,
this shifted minimum leads to an energy density
\cite{lyth,yana}.
\begin{equation}
\left(m_\phi Y_\phi\right)_{\rm TI} \approx 
\frac{c^4T_RV_0}{m_\phi^2M_P^2}\left(\frac{\phi_1}{M_P}\right)^2 \ ,
\label{TIb}
\end{equation}
where $c$ and $\phi_1$ are those  in the free
energy density (\ref{eq:effective-potential})
during thermal inflation.
We showed the resulting constraints on $\phi_1$ in Figure~3(b)
for the case that $c\sim 1$ over the entire moduli mass range,
and in Figure~3(c) for the more plausible case that
$c\approx m_{\phi}/m_{3/2}$ with $m_{3/2}=100\GeV$.

\section{Conclusion}

In this paper, we have discussed the possibility of light moduli having 
a mass far 
below the weak scale, and examined the cosmological bounds on the initial
moduli misalignment for the mass range $40\TeV\sim 10^{-26}\eV$.
A very light  moduli can arise as an  axion-like moduli in
string/$M$-theory with a large compactification radius,
$m_{\phi}\approx e^{-\pi {\rm Re}({\cal T})}m_{3/2}$
with ${\rm Re}({\cal T})=1\sim \frac{1}{\alpha_{GUT}}$.
Also generic moduli in gauge-mediated supersymmetry breaking models,
can have   a mass in the range $10\eV\sim1\GeV$.
We then studied the cosmological evolution of a generic
light modulus $\phi$ to
quantify its relic energy density which is determined
by the initial
misalignment $\delta\phi$.  
The initial misalignment $\delta\phi$ 
is set by  the modulus value $\phi_i$ at an initial time
$t_i$ with $\phi^{\prime}(t_i)=0$ 
and also by the dynamical parameter $c^2\phi_1$
in the moduli free energy density
(\ref{eq:effective-potential}) at $t\gtrsim t_i$.

For the case that there is no entropy production
after $\phi$-oscillation begins at $t\sim m_{\phi}^{-1}$
(but of course before $\phi$ decays),
the bounds on the misalignment $\delta\phi$
coming from various astronomical and cosmological  observations are shown
in Figure~1.  We then discussed how much such bounds can be relaxed by
a late entropy dumping which can be driven typically by e.g. 
out-of-equilibrium decay of another heavy
moduli [Figure~2] or by a late vacuum domination like thermal inflation
[Figure~3].
The bound on the initial misalignment turned out to be severe so that 
$\delta\phi\approx M_P$ is not allowed in most cases.
Thermal inflation appears to be efficiently  relax the bound,
however still $\delta\phi\approx M_P$ is allowed
only for a limited moduli mass range 
as shown in Figure 3.

\bigskip

{\bf Acknowledgments}:
KC and HBK thank KIAS for its hospitality during which this work
has been completed.
This work is supported by
KAIST Basic Science Research Program,
Basic Science Research Institutes Program, BSRI-97-2434,
and KOSEF through CTP of Seoul National University.


%
%

\newcommand{\te}[1]{$10^{#1}$}
\newcommand{\timeline}[3]{%
    \put {\boldmath#2} #3 at #1 0.8
    \plot #1 -13.0  #1 0.5 /
    \plot #1   1.2  #1 2.0 /
}

%
%

\begin{figure}
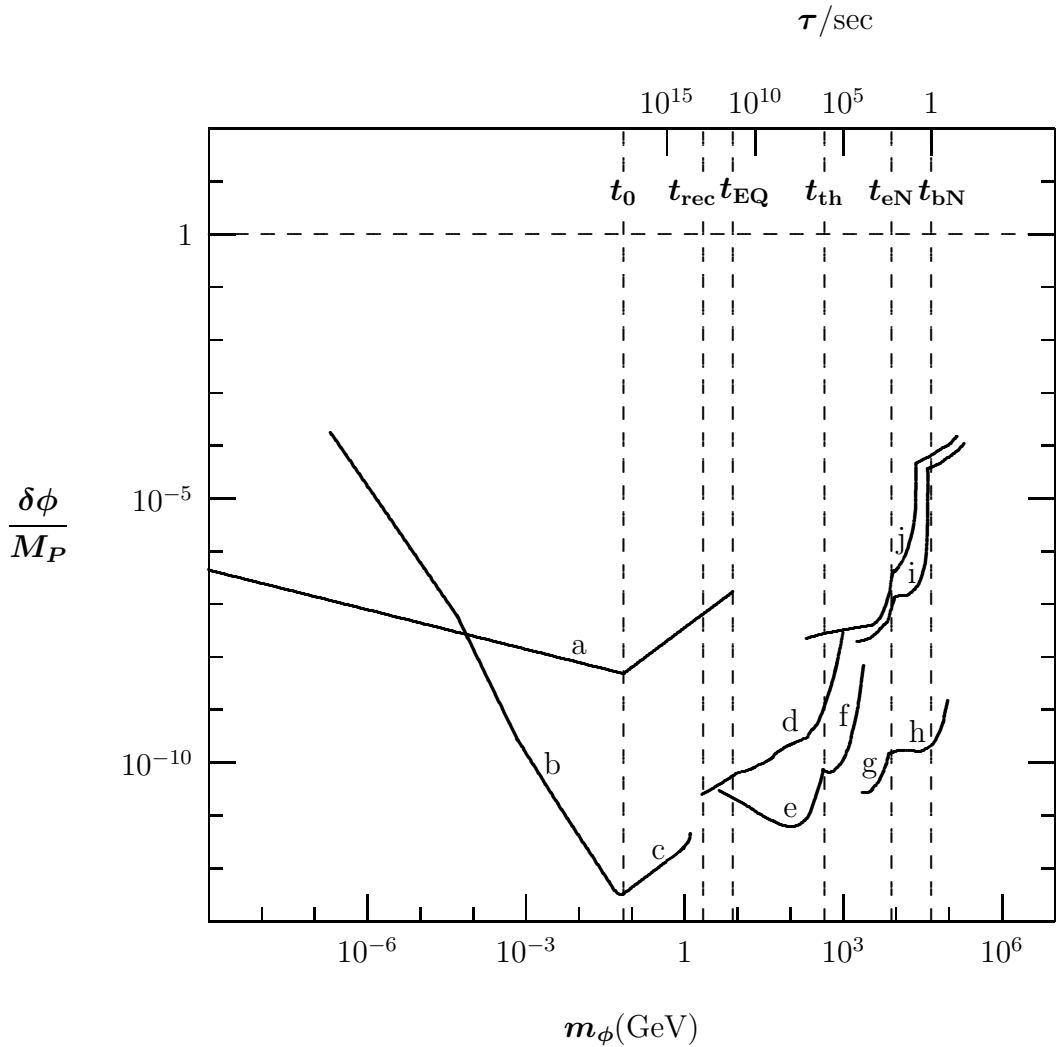


\caption{
Constraints on the initial misalignment of moduli.
We take $\Omega_0=1$, $h=0.7$ and the lifetime
$\tau=10^{14}\left(\frac{m_\phi}{1{\rm GeV}}\right)^{-3}$sec.
The line (a) comes from the lower bound on the age of the universe,
$\Omega_\phi h^2\le1$, which is unavoidable even for very light moduli.
The lines from (b) to (j) are for the case that moduli decay to photons
or charged particles:
(b) Recently reported X-ray background,
(c) Observed $\gamma$-ray background,
(d) Distorsion of CMBR, $\mu>8\times10^{-3}$,
(e) Photoproduction, $({\rm D}+{}^3{\rm He})/{\rm H}>10^{-4}$,
(f) Photodestruction, ${\rm D}/{\rm H}<10^{-5}$,
(g) Hadronic shower, $({\rm D}+{}^3{\rm He})/{\rm H}>10^{-4}$,
(h) Hadronic shower, $Y_p({}^4{\rm He})>0.25$,
(i) Entropy production, $({\rm D}+{}^3{\rm He})/{\rm H}>10^{-4}$,
(j) Entropy production, $Y_p({}^4{\rm He})>0.25$.
The line (a) can be straightforwardly 
extended  to the region
$m_{\phi} < 1\eV$.
}

$$\beginpicture
\setcoordinatesystem units <20pt,20pt> point at 0 0
\setplotarea x from -9.0 to 7.0, y from -13.0 to 2.0
\inboundscheckon
\linethickness 0.5pt
\axis bottom label {\boldmath$m_\phi$(GeV)}
      ticks in
            width <0.5pt> length <5.0pt>
            unlabeled quantity 17
            width <0.5pt> length <10.0pt>
            withvalues {\te{-6}} {\te{-3}} {1} {\te{3}} {\te{6}} /
            at -6 -3 0 3 6 /
/
\axis left label {\boldmath$\displaystyle\frac{\delta\phi}{M_P}$}
      ticks in
            width <0.5pt> length <5.0pt>
            unlabeled quantity 16
            width <0.5pt> length <10.0pt>
            withvalues {\te{-10}} {\te{-5}} {1} /
            at -10 -5 0 /
/
\axis top label {\makebox[0mm][l]{\hspace*{22mm}\boldmath$\tau$/sec}}
      ticks in
            width <0.5pt> length <10.0pt> 
            withvalues {\te{15}} {\te{10}} {\te{5}} {1} /
            at -0.3363 1.3363 3 4.6637 /
/
\axis right label {\phantom{\boldmath$\displaystyle\frac{\delta\phi}{M_P}$}}
      ticks in
            width <0.5pt> length <5.0pt>
            unlabeled quantity 16
            width <0.5pt> length <10.0pt>
            withvalues {} {} {} /
            at -10 -5 0 /
/
%
%
\setplotsymbol ({\makebox[0mm]{\tiny.}})
\setlinear
\setdashes
\plot -9.0 0.0  7.0 0.0 /
\timeline{-1.155}{$t_0$}{[c]}
\timeline{0.3527}{$t_{\rm rec}$}{[r] <5pt,0pt>}
\timeline{0.9107}{$t_{\rm EQ}$}{[l] <-5pt,0pt>}
\timeline{2.644}{$t_{\rm th}$}{[c]}
\timeline{3.912}{$t_{\rm eN}$}{[c]}
\timeline{4.6637}{$t_{\rm bN}$}{[l] <-5pt,0pt>}
\setplotsymbol (.)
\setsolid
\put {a} at -2.0 -7.8
\setlinear
\plot -9.000 -6.351 -1.155 -8.312 /
\plot -1.155 -8.312 0.9107 -6.763 /
\put {b} at -2.5 -10.0
\plot -6.6990 -3.7422 -4.3010 -7.2193 /
\plot -4.3010 -7.2197 -3.1549 -9.5692 /
\plot -3.1549 -9.5638 -2.3979 -10.7750 /
\plot -2.3979 -10.775 -1.3287 -12.4049 /
\setquadratic
\put {j} at 4.1     -5.8
\plot
   5.1500  -3.8127
   4.9977  -3.9783
   4.8724  -4.0389
   4.6440  -4.1921
   4.3798  -4.3231
/
\plot
   4.3798  -4.3231
   4.3798  -4.5859
   4.3753  -4.9132
   4.3619  -5.2383
   4.3216  -5.5567
   4.2455  -5.8662
   4.1604  -6.1077
   4.0395  -6.3074
   3.9724  -6.3760
   3.9276  -6.4436
   3.8828  -6.7609
   3.8067  -7.0047
   3.7216  -7.2199
   3.6365  -7.3234
   3.5380  -7.3974
/
\plot
   3.5380  -7.3974
   3.2962  -7.4354
   2.9783  -7.4873
   2.6156  -7.5609
   2.2977  -7.6456
/
\put {i} at
   4.3     -6.4
\plot
   5.2843  -3.9580
   5.1365  -4.1116
   4.9574  -4.2376
   4.7783  -4.3570
   4.5992  -4.4240
/
\plot
   4.5992  -4.4240
   4.5992  -4.9561
   4.5947  -5.4608
   4.5768  -5.9490
   4.5544  -6.2062
   4.5007  -6.4227
   4.4201  -6.6325
   4.2992  -6.7665
   4.1873  -6.8370
   4.0395  -6.8395
   3.9679  -6.8807
   3.9276  -7.0151
   3.8470  -7.3235
   3.7485  -7.4171
   3.6321  -7.5391
   3.4574  -7.6597
   3.2470  -7.7056
/
\put {h} at
   4.4     -9.4
\plot
   4.9773  -8.8187
   4.9367  -9.0134
   4.8920  -9.2286
   4.7822  -9.5084
   4.6725  -9.6697
   4.5099  -9.7638
   4.4286  -9.7866
   4.3554  -9.7791
   4.1603  -9.7573
   4.0220  -9.7766
   3.8676  -9.8296
/
\put {g} at
   3.5    -10.1
\plot
   3.8676  -9.8296
   3.8269  -9.9812
   3.7781 -10.1199
   3.7253 -10.2523
   3.6603 -10.3708
   3.5749 -10.4735
   3.5342 -10.5172
   3.4733 -10.5559
   3.3513 -10.5577
/
\put {f} at
   3.0     -9.1
\plot
   3.3838  -8.1508
   3.3513  -8.4122
   3.3229  -8.6477
   3.2985  -8.8302
   3.2578  -9.0411
   3.2172  -9.2465
   3.1684  -9.4446
   3.1155  -9.6578
   3.0383  -9.8649
   2.9651 -10.0029
   2.8513 -10.1201
   2.7375 -10.1832
   2.6237 -10.1171
/
\put {e} at
   2.0    -10.9
\plot
   2.6237 -10.1171
   2.5790 -10.3215
   2.5180 -10.5327
   2.4570 -10.7061
   2.3838 -10.9304
   2.2741 -11.0917
   2.1196 -11.1878
   1.9407 -11.1970
   1.7456 -11.1375
   1.5708 -11.0506
   1.3798 -10.9489
   1.1968 -10.8224
   0.9936 -10.7069
   0.8107 -10.6072
   0.6481 -10.5234
/
\put {d} at
   2.0     -9.2
\plot
   2.3192  -9.5183
   2.0799  -9.6230
   1.9773  -9.6693
   1.8919  -9.7097
   1.8166  -9.7680
   1.7278  -9.8228
   1.6628  -9.8991
   1.5739  -9.9592
   1.3278 -10.0827
   1.2423 -10.1231
   0.9859 -10.1978
   0.6474 -10.4165
   0.3192 -10.6018
/
\plot
   2.9961  -7.5435
   2.9004  -8.0338
   2.8423  -8.2763
   2.7842  -8.5085
   2.6953  -8.7794
   2.6337  -8.9748
   2.5654  -9.1274
   2.5107  -9.2628
   2.4355  -9.3571
   2.3910  -9.3974
   2.3192  -9.5183
/
\put {c} at
  -0.5    -11.7
\plot
   0.1097 -11.3322
   0.0960 -11.4365
   0.0745 -11.4894
   0.0433 -11.5340
  -0.0211 -11.6052
  -0.0816 -11.6687
  -0.3529 -11.8716
  -0.6886 -12.1283
  -1.0203 -12.3743
  -1.0711 -12.4198
  -1.1257 -12.4557
  -1.1765 -12.4867
  -1.2213 -12.4900
  -1.2682 -12.4638
  -1.3287 -12.4049
/
\endpicture$$

\end{figure}

%
%

\begin{figure}

\caption{
Relaxed bounds on the initial misalignment of light moduli
when there is an entropy production due to the heavy modulus decay
which yields the reheating temperature $T_{RH}=6$ MeV.
Again the line at $m_{\phi}\sim 1$ eV can be straightforwardly
extended to the region $m_{\phi}< 1\eV$.
}

$$\beginpicture
\setcoordinatesystem units <20pt,20pt> point at 0 0
\setplotarea x from -9.0 to 7.0, y from -8.0 to 2.0
\inboundscheckon
\linethickness 0.5pt
\axis bottom label {\boldmath$m_\phi$(GeV)}
      ticks in
            width <0.5pt> length <5.0pt>
            unlabeled quantity 17
            width <0.5pt> length <10.0pt>
            withvalues {\te{-6}} {\te{-3}} {1} {\te{3}} {\te{6}} /
            at -6 -3 0 3 6 /
/
\axis left label {\boldmath$\displaystyle\frac{\delta\phi}{M_P}$}
      ticks in
            width <0.5pt> length <5.0pt>
            unlabeled quantity 11
            width <0.5pt> length <10.0pt>
            withvalues {\te{-5}} {1} /
            at -5 0 /
/
\axis top label {\makebox[0mm][l]{\hspace*{22mm}\boldmath$\tau$(sec)}}
      ticks in
            width <0.5pt> length <10.0pt> 
            withvalues {\te{15}} {\te{10}} {\te{5}} {1} /
            at -0.3363 1.3363 3 4.6637 /
/
\axis right label {\phantom{\boldmath$\displaystyle\frac{\delta\phi}{M_P}$}}
      ticks in
            width <0.5pt> length <5.0pt>
            unlabeled quantity 11
            width <0.5pt> length <10.0pt>
            withvalues {} {} /
            at -5 0 /
/
\setplotsymbol ({\makebox[0mm]{\tiny.}})
\setlinear
\setdashes
\plot -9.0 0.0  7.0 0.0 /

\setplotsymbol (.)
\setsolid
\setlinear
\plot	-9.000 -3.114	-1.155 -3.114 /
\plot	-1.155 -3.114	0.9107 -1.048 /
\setlinear
\plot	-6.6990 -0.0691	-4.3010 -2.8086 /
\plot	-4.3010 -2.8090 -3.1549 -4.8719 /
\plot	-3.1549 -4.8665 -2.3979 -5.8885 /
\plot	-2.3979 -5.8885 -1.3287 -7.2511 /

\setquadratic
\plot
   5.1500   2.9608
   4.9977   2.7571
   4.8724   2.6652
   4.6440   2.4549
   4.3798   2.2578
/
\plot
   4.3798   2.2578
   4.3798   1.9950
   4.3753   1.6666
   4.3619   1.3382
   4.3216   1.0097
   4.2455   0.6812
   4.1604   0.4184
   4.0395   0.1885
   3.9724   0.1031
   3.9276   0.0243
   3.8828  -0.3042
   3.8067  -0.5670
   3.7216  -0.8035
   3.6365  -0.9283
   3.5380  -1.0269
/
\plot
   3.5380  -1.0269
   3.2962  -1.1254
   2.9783  -1.2567
   2.6156  -1.4210
   2.2977  -1.5852
/
%
\plot
   5.2843   2.8491
   5.1365   2.6585
   4.9574   2.4878
   4.7783   2.3236
   4.5992   2.2118
/
\plot
   4.5992   2.2118
   4.5992   1.6797
   4.5947   1.1739
   4.5768   0.6812
   4.5544   0.4184
   4.5007   0.1885
   4.4201  -0.0415
   4.2992  -0.2057
   4.1873  -0.3042
   4.0395  -0.3436
   3.9679  -0.4027
   3.9276  -0.5472
   3.8470  -0.8758
   3.7485  -0.9940
   3.6321  -1.1451
   3.4574  -1.3094
   3.2470  -1.4079
/
\plot
   4.9773  -2.0884
   4.9367  -2.2932
   4.8920  -2.5196
   4.7822  -2.8268
   4.6725  -3.0156
   4.5099  -3.1503
   4.4286  -3.1934
   4.3554  -3.2043
   4.1603  -3.2312
   4.0220  -3.2851
   3.8676  -3.3767
/
\plot
   3.8676  -3.3767
   3.8269  -3.5385
   3.7781  -3.6894
   3.7253  -3.8350
   3.6603  -3.9697
   3.5749  -4.0938
   3.5342  -4.1476
   3.4733  -4.2016
   3.3513  -4.2339
/
\plot
   3.3838  -1.8189
   3.3513  -2.0884
   3.3229  -2.3310
   3.2985  -2.5196
   3.2578  -2.7407
   3.2172  -2.9562
   3.1684  -3.1665
   3.1155  -3.3929
   3.0383  -3.6193
   2.9651  -3.7756
   2.8513  -3.9213
   2.7375  -4.0128
   2.6237  -3.9752
/
\plot
   2.6237  -3.9752
   2.5790  -4.1908
   2.5180  -4.4172
   2.4570  -4.6059
   2.3838  -4.8484
   2.2741  -5.0372
   2.1196  -5.1719
   1.9407  -5.2258
   1.7456  -5.2151
   1.5708  -5.1719
   1.3798  -5.1180
   1.1968  -5.0372
   0.9936  -4.9725
   0.8107  -4.9185
   0.6481  -4.8754
/
\plot
   2.3192  -3.4525
   2.0799  -3.6170
   1.9773  -3.6890
   1.8919  -3.7507
   1.8166  -3.8279
   1.7278  -3.9049
   1.6628  -3.9974
   1.5739  -4.0797
   1.3278  -4.2648
   1.2423  -4.3265
   0.9859  -4.4653
   0.6474  -4.7687
   0.3192  -5.0360
/
\plot
   2.9961  -1.3085
   2.9004  -1.8227
   2.8423  -2.0797
   2.7842  -2.3264
   2.6953  -2.6196
   2.6337  -2.8304
   2.5654  -3.0000
   2.5107  -3.1491
   2.4355  -3.2622
   2.3910  -3.3136
   2.3192  -3.4525
/
\plot
   0.1097  -5.8188
   0.0960  -5.9265
   0.0745  -5.9848
   0.0433  -6.0372
  -0.0211  -6.1245
  -0.0816  -6.2031
  -0.3529  -6.4738
  -0.6886  -6.8144
  -1.0203  -7.1434
  -1.0711  -7.2016
  -1.1257  -7.2511
  -1.1765  -7.2948
  -1.2213  -7.3093
  -1.2682  -7.2949
  -1.3287  -7.2511
/

\endpicture$$

\end{figure}

%
%

\begin{figure}
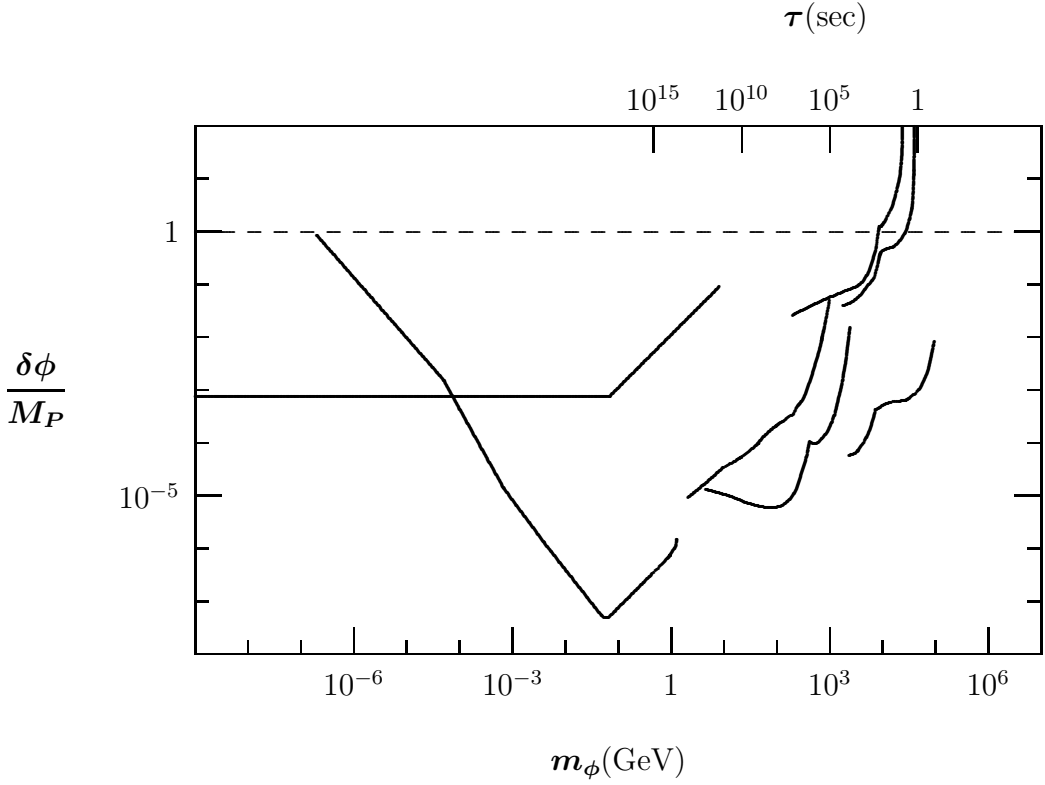


\caption{
Constraint on the misalignment of moduli when there exists thermal inflation
with $T_R=10{\rm MeV}$:
(a) Constraint on the initial misalignment of moduli,
(b) Constraint on the misalignment ($\delta\phi=\phi_1$) 
induced by the shifted minimum
during  thermal inflation when $c=1$ over the entire mass range,
(c) Constraint on the misalignment ($\delta\phi=\phi_1$)
induced by thermal inflation for $c=m_{\phi}/m_{3/2}$
with $m_{3/2}=100\GeV$.
}


$$\beginpicture
\setcoordinatesystem units <20pt,20pt> point at 0 0
\setplotarea x from -9.0 to 7.0, y from -6.0 to 2.0
\inboundscheckon
\linethickness 0.5pt
\axis bottom label {}
      ticks in
      width <0.5pt> length <5.0pt>
      unlabeled quantity 17
      width <0.5pt> length <10.0pt>
      at -6 -3 0 3 6 /
/
\axis left label {\boldmath\phantom{$\displaystyle\frac{\delta\phi}{M_P}$}}
      ticks in
      width <0.5pt> length <5.0pt>
      unlabeled quantity 9
      width <0.5pt> length <10.0pt>
      withvalues {\te{-5}} {1} /
      at -5 0 /
/
\axis top label {\makebox[0mm][l]{\hspace*{22mm}\boldmath$\tau$/sec}}
      ticks in
      width <0.5pt> length <10.0pt> 
      withvalues {\te{15}} {\te{10}} {\te{5}} {1} /
      at -0.3363 1.3363 3 4.6637 /
/
\axis right label {\phantom{\boldmath$\displaystyle\frac{\delta\phi}{M_P}$}}
      ticks in
      width <0.5pt> length <5.0pt>
      unlabeled quantity 9
      width <0.5pt> length <10.0pt>
      withvalues {} {} /
      at -5 0 /
/

\put {(a)} at -8 -5

\setplotsymbol ({\makebox[0mm]{\tiny.}})
\setlinear
\setdashes
\plot -9.0 0.0  7.0 0.0 /

%
%
\setplotsymbol (.)
\setsolid

\setlinear
\plot
  -9.000   -3.114
  -7.000   -3.114
/
\plot
  -7.000   -3.114
  -7.000    2.000
/
\plot
  -9.0000   2.7995
  -1.1550   0.8385
/
\plot
  -1.1550   0.8385
   0.9107   2.3875
/

\setlinear
\plot
  -6.6990   5.4083
  -4.3010   1.9312
/
\plot
  -4.3010   1.9308
  -3.1549  -0.4187
/
\plot
  -3.1549  -0.4133
  -2.3979  -1.6245
/
\plot
  -2.3979  -1.6245
  -1.3287  -3.2544
/

\setquadratic
\plot
   4.3798   4.8274
   4.3798   4.5646
   4.3753   4.2373
   4.3619   3.9122
   4.3216   3.5938
   4.2455   3.2843
   4.1604   3.0428
   4.0395   2.8431
   3.9724   2.7745
   3.9276   2.7069
   3.8828   2.3896
   3.8067   2.1458
   3.7216   1.9306
   3.6365   1.8271
   3.5380   1.7531
/
\plot
   3.5380   1.7531
   3.2962   1.7151
   2.9783   1.6632
   2.6156   1.5896
   2.2977   1.5049
/
\plot
   4.5992   4.7265
   4.5992   4.1944
   4.5947   3.6897
   4.5768   3.2015
   4.5544   2.9443
   4.5007   2.7278
   4.4201   2.5180
   4.2992   2.3840
   4.1873   2.3135
   4.0395   2.3110
   3.9679   2.2698
   3.9276   2.1354
   3.8470   1.8270
   3.7485   1.7334
   3.6321   1.6114
   3.4574   1.4908
   3.2470   1.4449
/
\plot
   4.9773   0.3318
   4.9367   0.1371
   4.8920  -0.0781
   4.7822  -0.3579
   4.6725  -0.5192
   4.5099  -0.6133
   4.4286  -0.6361
   4.3554  -0.6286
   4.1603  -0.6068
   4.0220  -0.6261
   3.8676  -0.6791
/
\plot
   3.8676  -0.6791
   3.8269  -0.8307
   3.7781  -0.9694
   3.7253  -1.1018
   3.6603  -1.2203
   3.5749  -1.3230
   3.5342  -1.3667
   3.4733  -1.4054
   3.3513  -1.4072
/
\plot
   3.3838   0.9997
   3.3513   0.7383
   3.3229   0.5028
   3.2985   0.3203
   3.2578   0.1094
   3.2172  -0.0960
   3.1684  -0.2941
   3.1155  -0.5073
   3.0383  -0.7144
   2.9651  -0.8524
   2.8513  -0.9696
   2.7375  -1.0327
   2.6237  -0.9666
/
\plot
   2.6237  -0.9666
   2.5790  -1.1710
   2.5180  -1.3822
   2.4570  -1.5556
   2.3838  -1.7799
   2.2741  -1.9412
   2.1196  -2.0373
   1.9407  -2.0465
   1.7456  -1.9870
   1.5708  -1.9001
   1.3798  -1.7984
   1.1968  -1.6719
   0.9936  -1.5564
   0.8107  -1.4567
   0.6481  -1.3729
/
\plot
   2.3192  -0.3678
   2.0799  -0.4725
   1.9773  -0.5188
   1.8919  -0.5592
   1.8166  -0.6175
   1.7278  -0.6723
   1.6628  -0.7486
   1.5739  -0.8087
   1.3278  -0.9322
   1.2423  -0.9726
   0.9859  -1.0473
   0.6474  -1.2660
   0.3192  -1.4513
/
\plot
   2.9961   1.6070
   2.9004   1.1167
   2.8423   0.8742
   2.7842   0.6420
   2.6953   0.3711
   2.6337   0.1757
   2.5654   0.0231
   2.5107  -0.1123
   2.4355  -0.2066
   2.3910  -0.2469
   2.3192  -0.3678
/
\plot
   0.1097  -2.1817
   0.0960  -2.2860
   0.0745  -2.3389
   0.0433  -2.3835
  -0.0211  -2.4547
  -0.0816  -2.5182
  -0.3529  -2.7211
  -0.6886  -2.9778
  -1.0203  -3.2238
  -1.0711  -3.2693
  -1.1257  -3.3052
  -1.1765  -3.3362
  -1.2213  -3.3395
  -1.2682  -3.3133
  -1.3287  -3.2544
/

%
%
\setplotsymbol (.)
\setdashes <2pt>

\setlinear
\plot
  -9.000   -3.114
  -4.155   -3.114
/
\plot
  -4.155   -3.114
  -4.155    2.000
/

\plot
  -3.1549   2.3252
  -2.3979   1.1140
/
\plot
  -2.3979   1.1140
  -1.3287  -0.5159
/

\setquadratic
\plot
   3.8676   2.0594
   3.8269   1.9078
   3.7781   1.7691
   3.7253   1.6367
   3.6603   1.5182
   3.5749   1.4155
   3.5342   1.3718
   3.4733   1.3331
   3.3513   1.3313
/
\plot
   3.3838   3.7382
   3.3513   3.4768
   3.3229   3.2413
   3.2985   3.0588
   3.2578   2.8479
   3.2172   2.6425
   3.1684   2.4444
   3.1155   2.2312
   3.0383   2.0241
   2.9651   1.8861
   2.8513   1.7689
   2.7375   1.7058
   2.6237   1.7719
/
\plot
   2.6237   1.7719
   2.5790   1.5675
   2.5180   1.3563
   2.4570   1.1829
   2.3838   0.9586
   2.2741   0.7973
   2.1196   0.7012
   1.9407   0.6920
   1.7456   0.7515
   1.5708   0.8384
   1.3798   0.9401
   1.1968   1.0666
   0.9936   1.1821
   0.8107   1.2818
   0.6481   1.3656
/
\plot
   2.3192   2.3707
   2.0799   2.2660
   1.9773   2.2197
   1.8919   2.1793
   1.8166   2.1210
   1.7278   2.0662
   1.6628   1.9899
   1.5739   1.9298
   1.3278   1.8063
   1.2423   1.7659
   0.9859   1.6912
   0.6474   1.4725
   0.3192   1.2872
/
\plot
   0.1097   0.5568
   0.0960   0.4525
   0.0745   0.3996
   0.0433   0.3550
  -0.0211   0.2838
  -0.0816   0.2203
  -0.3529   0.0174
  -0.6886  -0.2393
  -1.0203  -0.4853
  -1.0711  -0.5308
  -1.1257  -0.5667
  -1.1765  -0.5977
  -1.2213  -0.6010
  -1.2682  -0.5748
  -1.3287  -0.5159
/

%
%
\setplotsymbol (.)
\setdots <2pt>
\setlinear
\plot
  -9.000   -3.114
  -4.000   -3.114
/
\plot
  -4.0050 -3.1140
  -3.1549 -4.8719
/
\plot
  -3.1549 -4.8665
  -3.0000 -5.0756
/
\plot
  -3.0000 -5.0756
  -3.0000  2.0000
/

\setlinear
\plot
  -2.3979   3.1770
  -1.3287   1.5471
/

\setquadratic
\plot
   0.1097   2.6198
   0.0960   2.5155
   0.0745   2.4626
   0.0433   2.4180
  -0.0211   2.3468
  -0.0816   2.2833
  -0.3529   2.0804
  -0.6886   1.8237
  -1.0203   1.5777
  -1.0711   1.5322
  -1.1257   1.4963
  -1.1765   1.4653
  -1.2213   1.4620
  -1.2682   1.4882
  -1.3287   1.5471
/


\endpicture$$

\kern-20mm


$$\beginpicture
\setcoordinatesystem units <20pt,20pt> point at 0 0
\setplotarea x from -9.0 to 7.0, y from -7.0 to 2.0
\inboundscheckon
\linethickness 0.5pt
\axis bottom label {} 
      ticks in
      width <0.5pt> length <5.0pt>
      unlabeled quantity 17
      width <0.5pt> length <10.0pt>
      at -6 -3 0 3 6 /
/
\axis left label {\boldmath$\displaystyle\frac{\delta\phi}{M_P}$}
      ticks in
      width <0.5pt> length <5.0pt>
      unlabeled quantity 10
      width <0.5pt> length <10.0pt>
      withvalues {\te{-5}} {1} /
      at -5 0 /
/
\axis top label {}
      ticks in
      width <0.5pt> length <10.0pt> 
      at -0.3363 1.3363 3 4.6637 /
/
\axis right label {\phantom{\boldmath$\displaystyle\frac{\delta\phi}{M_P}$}}
      ticks in
      width <0.5pt> length <5.0pt>
      unlabeled quantity 10
      width <0.5pt> length <10.0pt>
      withvalues {} {} /
      at -5 0 /
/

\put {(b)} at -8 -6

\setplotsymbol ({\tiny.})
\setlinear
\setdashes
\plot -9.0 0.0  7.0 0.0 /

%
%
\setplotsymbol (.)
\setsolid
\plot
  -9.000   -3.114
  -7.000   -3.114
/
\plot
  -7.000   -3.114
  -7.000    1.288
/

\setlinear
\plot
  -7.0000   1.2880
  -1.1550   7.1333
/
\plot
  -1.1550   7.1333
   0.9107  11.2644
/

\setlinear
\plot
  -6.6990   4.7730
  -4.3010   4.2934
/
\plot
  -4.3010   4.2931
  -3.1549   3.3762
/
\plot
  -3.1549   3.3816
  -2.3979   3.1166
/
\plot
  -2.3979   3.1166
  -1.3287   2.8232
/

\setsolid
\setquadratic

%
%
\setplotsymbol (.)
\setdashes <2pt>

\setlinear
\plot
  -9.000   -3.114
  -4.155   -3.114
/
\plot
  -4.155   -3.114
  -4.155   -2.761
/

\setlinear
\plot
  -4.1555  -2.7612
  -3.1549  -3.5613
/
\plot
  -3.1549  -3.5559
  -2.3979  -3.8209
/
\plot
  -2.3979  -3.8209
  -1.3287  -4.1143
/

\setquadratic
\plot
   2.6237   3.1140
   2.5790   2.8537
   2.5180   2.5663
   2.4570   2.3167
   2.3838   2.0008
   2.2741   1.7024
   2.1196   1.4132
   1.9407   1.1804
   1.7456   0.9960
   1.5708   0.8644
   1.3798   0.7274
   1.1968   0.6251
   0.9936   0.4866
   0.8107   0.3577
   0.6481   0.2382
/
\plot
   2.3192   3.3322
   2.0799   2.9284
   1.9773   2.7538
   1.8919   2.6067
   1.8166   2.4543
   1.7278   2.2885
   1.6628   2.1309
   1.5739   1.9597
   1.3278   1.5286
   1.2423   1.3813
   0.9859   0.9861
   0.6474   0.3443
   0.3192  -0.2513
/
\plot
   0.1097  -1.2436
   0.0960  -1.3650
   0.0745  -1.4448
   0.0433  -1.5284
  -0.0211  -1.6801
  -0.0816  -1.8192
  -0.3529  -2.3612
  -0.6886  -3.0375
  -1.0203  -3.6982
  -1.0711  -3.8072
  -1.1257  -3.9113
  -1.1765  -4.0058
  -1.2213  -4.0651
  -1.2682  -4.0976
  -1.3287  -4.1143
/

%
%
\setplotsymbol (.)
\setdots <2pt>

\setlinear

\setlinear
\plot
  -9.000   -3.114
  -4.000   -3.114
/
\plot
  -4.0050 -3.1140
  -3.1549 -4.8719
/
\plot
  -3.1549 -4.8665
  -3.0000 -5.0756
/
\plot
  -3.0000 -5.0756
  -3.0000 -6.3230
/
\plot
  -3.0000  -6.3230
  -2.3979  -6.5338
/
\plot
  -2.3979  -6.5338
  -1.3287  -6.8272
/

\setquadratic
\plot
   3.8676   2.2435
   3.8269   2.0410
   3.7781   1.8413
   3.7253   1.6429
   3.6603   1.4432
   3.5749   1.2337
   3.5342   1.1392
   3.4733   1.0243
   3.3513   0.8700
/
\plot
   3.3838   3.3176
   3.3513   3.0155
   3.3229   2.7445
   3.2985   2.5315
   3.2578   2.2698
   3.2172   2.0136
   3.1684   1.7545
   3.1155   1.4752
   3.0383   1.1716
   2.9651   0.9421
   2.8513   0.6826
   2.7375   0.4773
   2.6237   0.4011
/
\plot
   2.6237   0.4011
   2.5790   0.1409
   2.5180  -0.1466
   2.4570  -0.3962
   2.3838  -0.7120
   2.2741  -1.0105
   2.1196  -1.2997
   1.9407  -1.5325
   1.7456  -1.7169
   1.5708  -1.8485
   1.3798  -1.9855
   1.1968  -2.0878
   0.9936  -2.2263
   0.8107  -2.3552
   0.6481  -2.4747
/
\plot
   2.3192   0.6193
   2.0799   0.2155
   1.9773   0.0409
   1.8919  -0.1062
   1.8166  -0.2586
   1.7278  -0.4244
   1.6628  -0.5820
   1.5739  -0.7532
   1.3278  -1.1844
   1.2423  -1.3316
   0.9859  -1.7268
   0.6474  -2.3687
   0.3192  -2.9642
/
\plot
   2.9961   3.4402
   2.9004   2.8303
   2.8423   2.5152
   2.7842   2.2104
   2.6953   1.8283
   2.6337   1.5559
   2.5654   1.3179
   2.5107   1.1142
   2.4355   0.9259
   2.3910   0.8299
   2.3192   0.6193
/
\plot
   0.1097  -3.9565
   0.0960  -4.0779
   0.0745  -4.1577
   0.0433  -4.2413
  -0.0211  -4.3930
  -0.0816  -4.5321
  -0.3529  -5.0741
  -0.6886  -5.7504
  -1.0203  -6.4111
  -1.0711  -6.5201
  -1.1257  -6.6242
  -1.1765  -6.7187
  -1.2213  -6.7780
  -1.2682  -6.8105
  -1.3287  -6.8272
/


\endpicture$$

\kern-20mm


$$\beginpicture
\setcoordinatesystem units <20pt,20pt> point at 0 0
\setplotarea x from -9.0 to 7.0, y from -6.0 to 2.0
\inboundscheckon
\linethickness 0.5pt
\axis bottom label {\boldmath$m_\phi$/GeV}
      ticks in
      width <0.5pt> length <5.0pt>
      unlabeled quantity 17
      width <0.5pt> length <10.0pt>
      withvalues {\te{-6}} {\te{-3}} {1} {\te{3}} {\te{6}} /
      at -6 -3 0 3 6 /
/
\axis left label {\boldmath\phantom{$\displaystyle\frac{\delta\phi}{M_P}$}}
      ticks in
      width <0.5pt> length <5.0pt>
      unlabeled quantity 9
      width <0.5pt> length <10.0pt>
      withvalues {\te{-5}} {1} /
      at -5 0 /
/
\axis top label {}
      ticks in
      width <0.5pt> length <10.0pt> 
      at -0.3363 1.3363 3 4.6637 /
/
\axis right label {\phantom{\boldmath$\displaystyle\frac{\delta\phi}{M_P}$}}
      ticks in
      width <0.5pt> length <5.0pt>
      unlabeled quantity 9
      width <0.5pt> length <10.0pt>
      withvalues {} {} /
      at -5 0 /
/

\put {(c)} at -8 -5

\setplotsymbol ({\tiny.})
\setlinear
\setdashes
\plot -9.0 0.0  7.0 0.0 /

%
%
\setplotsymbol (.)
\setsolid

\setlinear
\plot -9.000   -3.114 -7.000   -3.114 /
\plot -7.000   -3.114 -7.000    2.000 /


\setquadratic

%
%
\setplotsymbol (.)
\setdashes <2pt>

\setlinear
\plot -9.000   -3.114 -4.155   -3.114 /
\plot -4.155   -3.114 -4.155    2.000 /


\setquadratic
\plot
   2.6237   1.8666
   2.5790   1.6957
   2.5180   1.5303
   2.4570   1.4027
   2.3838   1.2332
   2.2741   1.1542
   2.1196   1.1740
   1.9407   1.2990
   1.7456   1.5048
   1.5708   1.7228
   1.3798   1.9678
   1.1968   2.2315
   0.9936   2.4994
   0.8107   2.7363
   0.6481   2.9420
/
\plot
   2.3192   2.6938
   2.0799   2.7686
   1.9773   2.7992
   1.8919   2.8229
   1.8166   2.8211
   1.7278   2.8329
   1.6628   2.8053
   1.5739   2.8119
   1.3278   2.8730
   1.2423   2.8967
   0.9859   3.0143
   0.6474   3.0495
   0.3192   3.1103
/
\plot
   0.1097   2.5370
   0.0960   2.4430
   0.0745   2.4062
   0.0433   2.3850
  -0.0211   2.3621
  -0.0816   2.3440
  -0.3529   2.3446
  -0.6886   2.3397
  -1.0203   2.3424
  -1.0711   2.3350
  -1.1257   2.3401
  -1.1765   2.3472
  -1.2213   2.3775
  -1.2682   2.4388
  -1.3287   2.5431
/

%
%
\setplotsymbol (.)
\setdots <2pt>

\setlinear
\plot -9.000   -3.114 -4.000   -3.114 /
\plot -4.0050 -3.1140 -3.1549 -4.8719 /
\plot -3.1549 -4.8665 -3.0000 -5.0756 /
\plot -3.0000 -5.0756 -3.0000  2.0000 /
\plot
  -3.0000   3.6770
  -2.3979   2.2620
/
\plot
  -2.3979   2.2620
  -1.3287  -0.1698
/

\setquadratic
\plot
   3.8676  -1.4917
   3.8269  -1.6128
   3.7781  -1.7149
   3.7253  -1.8077
   3.6603  -1.8774
   3.5749  -1.9161
   3.5342  -1.9292
   3.4733  -1.9223
   3.3513  -1.8326
/
\plot
   3.3838   0.5500
   3.3513   0.3129
   3.3229   0.0987
   3.2985  -0.0655
   3.2578  -0.2458
   3.2172  -0.4208
   3.1684  -0.5823
   3.1155  -0.7558
   3.0383  -0.9050
   2.9651  -0.9881
   2.8513  -1.0200
   2.7375  -0.9977
   2.6237  -0.8463
/
\plot
   2.6237  -0.8463
   2.5790  -1.0171
   2.5180  -1.1826
   2.4570  -1.3102
   2.3838  -1.4796
   2.2741  -1.5587
   2.1196  -1.5389
   1.9407  -1.4139
   1.7456  -1.2081
   1.5708  -0.9901
   1.3798  -0.7451
   1.1968  -0.4814
   0.9936  -0.2135
   0.8107   0.0234
   0.6481   0.2291
/
\plot
   2.3192  -0.0191
   2.0799   0.0557
   1.9773   0.0863
   1.8919   0.1100
   1.8166   0.1082
   1.7278   0.1200
   1.6628   0.0924
   1.5739   0.0990
   1.3278   0.1600
   1.2423   0.1838
   0.9859   0.3014
   0.6474   0.3365
   0.3192   0.3974
/
\plot
   2.9961   1.4480
   2.9004   1.0295
   2.8423   0.8306
   2.7842   0.6420
   2.6953   0.4377
   2.6337   0.2885
   2.5654   0.1871
   2.5107   0.0928
   2.4355   0.0549
   2.3910   0.0479
   2.3192  -0.0191
/
\plot
   0.1097  -0.1759
   0.0960  -0.2699
   0.0745  -0.3067
   0.0433  -0.3279
  -0.0211  -0.3508
  -0.0816  -0.3689
  -0.3529  -0.3683
  -0.6886  -0.3732
  -1.0203  -0.3705
  -1.0711  -0.3779
  -1.1257  -0.3728
  -1.1765  -0.3657
  -1.2213  -0.3354
  -1.2682  -0.2741
  -1.3287  -0.1698
/

\setlinear
\setsolid        \plot 3.0 -4.0 4.5 -4.0 /
\setdashes <2pt> \plot 3.0 -4.5 4.5 -4.5 /
\setdots <2pt>   \plot 3.0 -5.0 4.5 -5.0 /
\put {$n=1$} <5pt,0pt> [l] at 4.5 -4.0
\put {$n=2$} <5pt,0pt> [l] at 4.5 -4.5
\put {$n=3$} <5pt,0pt> [l] at 4.5 -5.0

\endpicture$$

\end{figure}

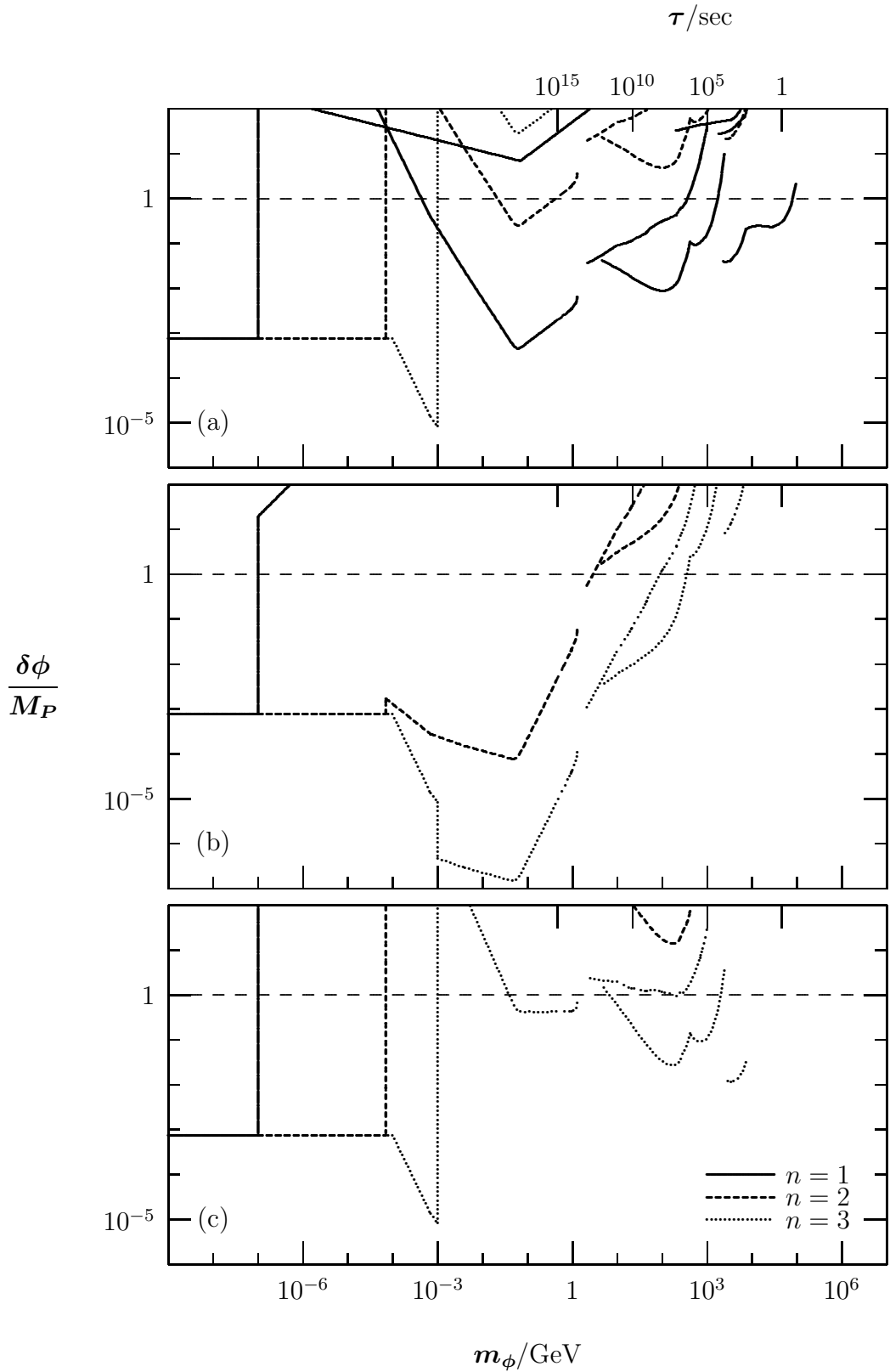
\end{document}